# Crystallization kinetics of (S)-4'-(1-methylheptyloxycarbonyl)biphenyl-4-yl 4-[4-(2,2,3,3,4,4,4-heptafluorobutoxy)but-1-oxy]-2-fluorobenzoate

Aleksandra Deptuch[1,*], Sebastian Lalik[2], Małgorzata Jasiurkowska-Delaporte[1], Magdalena Urbańska[3], Monika Marzec[2]

[1] Institute of Nuclear Physics Polish Academy of Sciences, Radzikowskiego 152, PL-31342 Kraków, Poland
[2] Institute of Physics, Jagiellonian University, Łojasiewicza 11, PL-30348 Kraków, Poland
[3] Institute of Chemistry, Military University of Technology, Kaliskiego 2, PL-00908 Warsaw, Poland
[*]corresponding author, aleksandra.deptuch@ifj.edu.pl

**Abstract**

Chiral liquid crystalline compounds belonging to the homologous series of (S)-4'-(1-methylheptyloxycarbonyl)biphenyl-4-yl 4-[m-(2,2,3,3,4,4,4-heptafluorobutoxy)alk-1-oxy]-2-fluorobenzoates show various behavior on cooling depending on the length of the $C_mH_{2m}$ chain. The homologue with m = 2 crystallizes, while for m = 5, 6, 7, and presumably also for m = 3, the glass of the anticlinic smectic $C_A$* phase is formed. The previous results for m = 4 suggest that this homologue may also be a glassformer. This paper presents the study of the crystallization kinetics for the compound with m = 4 in isothermal conditions (by polarizing optical microscopy) and for the 5-40 K/min cooling rates (by differential scanning calorimetry). Microscopic observations enable estimation of the energy barrier for nucleation, which equals 409 kJ/mol. The threshold cooling rate necessary for complete vitrification of the smectic $C_A$* phase, obtained by extrapolating the enthalpy change during crystallization to zero, is equal to 81 K/min or 64 K/min for the linear and parabolic fit, respectively. The structural studies by X-ray diffraction show that crystal phases have lamellar structures both in the pristine sample and after crystallization from the melt but with different layer spacing. A weak relaxation process is detected in the sample after melt crystallization, revealing the presence of the conformational disorder. The dynamic glass transition temperature of the SmC$_A$* phase, estimated from the relaxation time of the $P_H$ process (as the α-relaxation time could not be registered in a wide enough temperature range), is 244 K.

## 1. Introduction

(S)-4'-(1-methylheptyloxycarbonyl)biphenyl-4-yl 4-[m-(2,2,3,3,4,4,4-heptafluorobutoxy)alk-1-oxy]-2-fluorobenzoates (abbreviated as 3FmHPhF6, Figure 1) are the homologous series of chiral mesogenic compounds which form various smectic phases, including the paraelectric SmA*, ferroelectric SmC* and antiferroelectric SmC$_A$* [1-7]. The 3FmHPhF6 series and other smectogenic compounds with similar molecular structures are used as components of the orthoconic antiferroelectric mixtures (with a tilt angle ≈ 45°) [1,2,8-12], which have suitable properties for application in liquid crystal displays [13,14]. For use in displays, one prefers a liquid crystalline material which does not form a crystal phase neither on cooling nor upon heating (cold crystallization), and such property can be obtained for mixtures [12]. The pure 3FmHPhF6 compounds, even when they form the glass of the smectic phase instead of undergoing crystallization, exhibit the crystal phase after subsequent heating from the vitrified state [5-7]. However, materials with such properties also have potential applications, e.g. for energy storage [15-18]. The length of the $C_mH_{2m}$ chain in the 3FmHPhF6 molecules impacts the sequence of the smectic phases, electro-optic properties, helical pitch [1-3], and their glassforming ability. The homologues with m = 5, 6, 7 undergo the vitrification of the SmC$_A$* phase for the rather low cooling rate of 3 K/min and crystallize on heating [5-7]. For m = 3, the glass transition is not confirmed, but the differential scanning calorimetry (DSC) results for the cooling/heating rate equal to 2.5 K/min published in Ref. [4] indicate that this homologue is probably a glassformer, because the formation of the crystal phase occurs only partially during cooling and is completed during subsequent heating. The homologue with m = 2 crystallizes on cooling and its vitrification was not observed [6]. This study is dedicated to the 3FmHPhF6 homologue with m = 4, which has intermediate properties. The phase sequence of 3F4HPhF6 on heating is Cr (328.1 K) SmC$_A$* (360.7 K) SmC* (380.2 K) Iso [2]. The DSC results from the supplementary material of Ref. [6] contain an exothermic anomaly on cooling with 3-20 K/min rates, interpreted as crystallization. However, for 20 K/min, the size of this anomaly indicates that crystallization is only partial, and at lower temperatures, the sample is a mixture of a crystal and the vitrified SmC$_A$* phase. To check this hypothesis and to have a better image of the potential glassforming properties of 3F4HPhF6, its crystallization process in isothermal and non-isothermal conditions is investigated using polarizing optical microscopy (POM) and differential scanning calorimetry (DSC), respectively. The crystal phases of 3F4HPhF6 are additionally studied by X-ray diffraction (XRD) and broadband dielectric spectroscopy (BDS) methods.

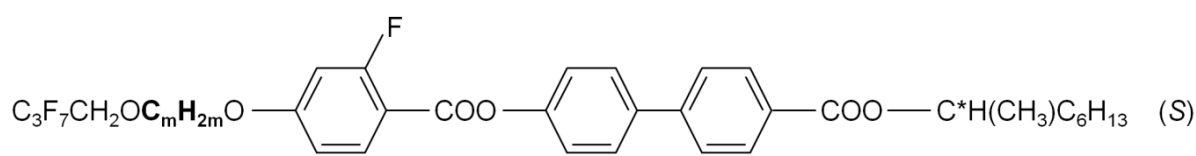

Figure 1. The general formula of the 3FmHPhF6 homologous series. This study presents results for m = 4.

## 2. Experimental details

The synthetic route of (S)-4'-(1-methylheptyloxycarbonyl)biphenyl-4-yl 4-[m-(2,2,3,3,4,4,4-heptafluorobutoxy)alk-1-oxy]-2-fluorobenzoates is presented in [1,2].

The POM observations were carried out using the Leica DM2700 P microscope equipped with the Linkam cooling/heating stage for a sample placed between two microscopic slides without aligning layers. The textures were taken in the configuration of crossed polarizes. Each texture covers an area of 1243 × 933 μm$^2$. The textures were analyzed numerically in the TOApy program [19], which converts textures into grayscale images and then calculates an average luminance of all pixels in the 0-255 scale. Additionally, ImageJ [20] and OriginPro were used to analyze crystallization kinetics.

The DSC measurements in the 213-403 K range were performed with the cooling/heating rates of 5-40 K/min using the PerkinElmer DSC 8000 calorimeter for a sample weighting 4.20 mg pressed within the aluminum pan of 30 μl volume. The calibration was based on indium and water's melting points and melting enthalpies. The data analysis was performed with the PerkinElmer software and OriginPro.

The XRD patterns were collected with the X'Pert PRO (PANalytical) diffractometer equipped with the TTK 450 (Anton Paar) cooling/heating stage for a flat polycrystalline sample in the Bragg-Brentano geometry. The measurements were done in the 2θ range of 2-30° with the CuKα radiation. The analysis of diffraction patterns was done in OriginPro.

The BDS measurements were performed using the Novocontrol Technologies spectrometer for a sample with a thickness of 80 μm placed between two gold electrodes without any aligning layers. The dielectric spectra were collected in the 0.1-10$^7$ Hz frequency range. The analysis was done using OriginPro.

## 3. Results and discussion
### *3.1. Phase sequence and isothermal crystallization investigated by POM observations*

The textures collected for the high cooling/heating rate equal to 30 K/min reveal four phases of 3F4HPhF6 in the temperature range of 193-398 K: isotropic liquid (Iso), SmC*, SmC$_A$* and a crystal phase (Figure 2). The crystallization process, visible at 270 K, is unusual because the growth of crystallites is not observed. Instead, the texture of the crystal phase resembles the fan-shaped texture of the SmC$_A$* phase, and it only has a different color. During heating, there are subtle changes in the brightness and color of a crystal's texture, which are visible also in the results of numerical analysis. The crystal → SmC$_A$* transition is observed at 324 K. Note that the crystallization and melting do not change the arrangement of defects in the fan-shaped texture of the SmC$_A$* phase – compare Figures 2c and 2h. The similar textures of the smectic and crystal phases suggest that the latter has some degree of disorder [21], the crystal structure resembles the lamellar order of the smectic phase [22], or the size of crystallites is very small.

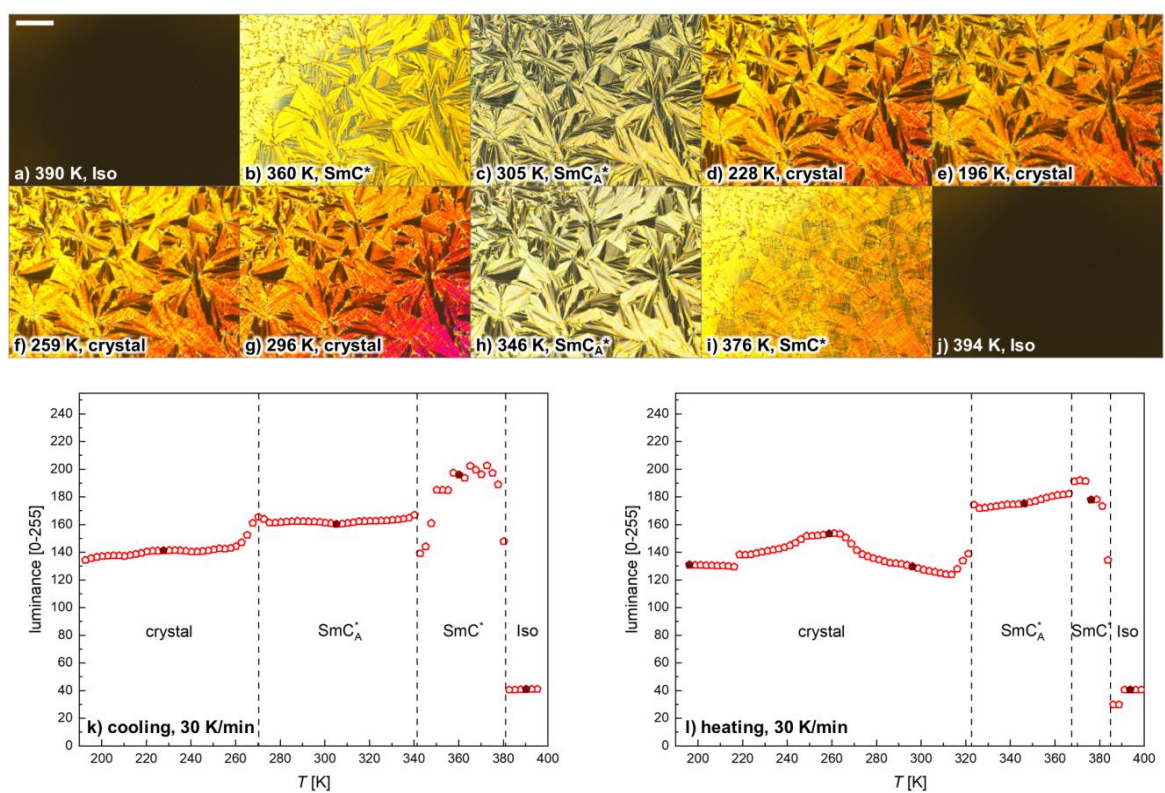

Figure 2. POM textures collected on cooling (a-d) and subsequent heating (e-j) with the 30 K/min rate and luminance vs. temperature plots for cooling (k) and heating (l) obtained by numerical analysis of the textures. The scale bar in (a) corresponds to 200 μm and applies to all the textures (a-j). The solid points in (k,l) indicate the textures presented in (a-j).

When the sample is cooled at 30 K/min to 283 K and kept at this temperature, the crystallization occurs in the same way as during cooling with the 30 K/min rate: only with the change in the color of texture (Figure 3a). The time necessary for a total SmC$_A$* → crystal transition is 90 s. Meanwhile, when the sample is cooled at 30 K/min to a higher temperature, 293 K, the typical crystallization with visible growing crystals is observed (Figure 3b), and the total crystallization time is almost 16 times longer, ca. 1410 s. The isothermal observations for intermediate temperatures (Figure S1 in Supplementary Materials) show that the amount of nuclei increases very quickly with the lowering of temperature. For 285 K, the crystallization resembles the one at 283 K, while for 289-292 K, the particular crystallites are visible, as for 293 K. The intermediate case is crystallization at 287 K, where the amount of nuclei is much larger than at higher temperatures, but they are still distinguishable.

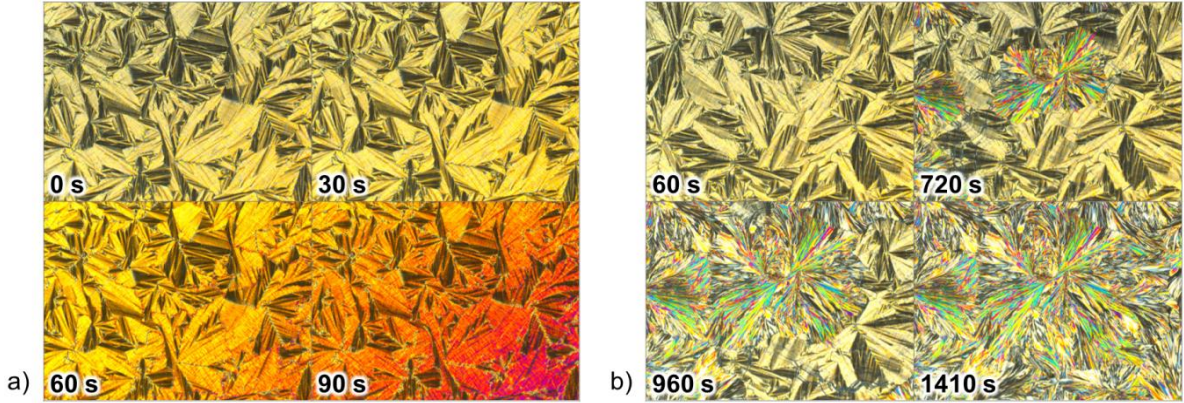

Figure 3. POM textures registered during isothermal crystallization in 283 K (a) and 293 K (b) after cooling with the 30 K/min rate from the isotropic liquid phase.

For $T_{cr}$ = 289-293 K, it is possible to obtain the time dependence of crystallization degree $X$ as the fractional area of textures covered by the crystal phase [23,24] (Figure 4a) and to count the amount of growing nuclei $N$ (Figure 4b). The uncertainty of $X$ was assumed to be 0.01. The experimental $X(t)$ dependence can be described by the Avrami model [25-27]:

$$X(t) = 1 - \exp\left(\left(\frac{t-t_0}{\tau_{cr}}\right)^n\right). \quad (1)$$

The initialization time $t_0$, characteristic time of crystallization $\tau_{cr}$ and Avrami parameter $n$ were obtained by fitting Equation (1) to the $X(t)$ plots (Figure 4a) and presented in Table 1. For 289-292 K, the $t_0$ values were around zero and with significant uncertainty, thus, they were fixed to zero, while for 293 K, the crystallization started after 5 min. The dimensionality of crystal growth affects the Avrami parameter $n$ [25-27]. For 293 K, $n$ = 2.7, which is within the range of the 2-dimensional growth. For 291 K and 292 K, $n$ = 3.1 and 4.0, corresponds to the isotropic crystal growth. Finally, for 289 K and 290 K, $n$ = 4.2 and 4.6, indicates the contribution of the sheaf-like growth. The $\tau_{cr}$ values are larger for higher $T_{cr}$ in the 289-292 K range and show a linear dependence in the activation plot (inset in Figure 4a). The negative effective activation energy, −126(10) kJ/mol, is not an energy barrier [28]. Instead, it indicates that the nucleation rate is the main factor that controls the crystallization kinetics. This is the effect of decreasing difference in free energy $\Delta G$ between the SmC$_A$* and crystal phases, equal approximately to $\Delta S_m(T_m - T)$, where $\Delta S_m$ and $T_m$ are the melting entropy and temperature, respectively [29]. Consequently, the maximal amount of nuclei $N_{max}$ decreases when $T_{cr}$ is higher, and crystallization occurs slower. The diffusion of molecules and crystal growth occur faster at higher temperatures, but for the melt crystallization of 3F4HPhF6 below 293 K, the contribution of nucleation rate is prevalent. The plot of $\ln N_{max}$ vs. $1000/T_{cr}$ (where the uncertainty of $N_{max}$ was assumed as 10% of $N_{max}$, rounded up to the integer) shows also an approximately linear dependence and the energy barrier for nucleation determined from its slope is equal to 409(48) kJ/mol (inset in Figure 4b). Based on the $N_{max}$, it is possible to estimate the average size of crystallites $D \approx \sqrt{A/N_{max}}$, where $A$ = 1243 × 933 μm² is the area of the texture. The crystallites grown at temperatures of 289-293 K have an average size of hundreds of micrometers, decreasing for lower $T_{cr}$. By extrapolation of the Arrhenius dependence from the inset in Figure 4b to 283 K, one can estimate the average size of crystallites equal to 18(5) μm. It is ~2% of the shorter frame of the observed area (933 μm). At 270 K, where the non-isothermal crystallization during cooling at 30 K/min is observed, the estimated size is only 0.3(2) μm. It explains why the defects of alignment present in the SmC$_A$* phase are preserved after crystallization, which occurs below 287 K – they are much larger than the crystallites.

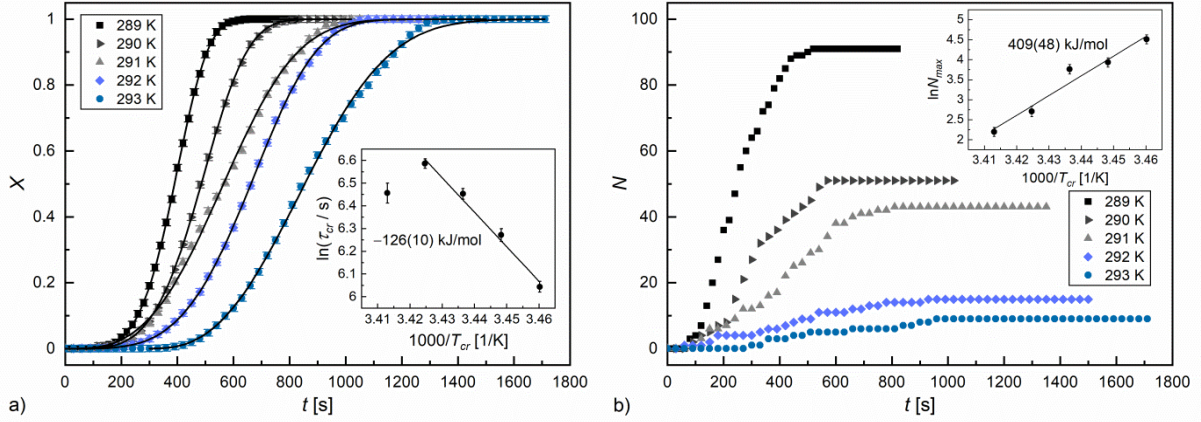

Figure 4. Crystallization degree (a) and amount of observed nuclei (b) during isothermal crystallization of 3F4HPhF6. The inset shows the activation plots for characteristic crystallization time (a) and maximal amount of nuclei (b).

Table 1. Parameters describing the isothermal crystallization of 3F4HPhF6 at different temperatures $T_{cr}$, determined from POM textures: initialization time $t_0$, characteristic time $\tau_{cr}$, Avrami parameter $n$, maximal amount of growing nuclei $N_{max}$ and average size of nuclei $D$.

| $T_{cr}$ [K] | $t_0$ [s] | $\tau_{cr}$ [s] | $n$ | $N_{max}$ | $D$ [μm] |
|---|---|---|---|---|---|
| 289 | 0 | 421(10) | 4.6(1) | 91(10) | 113(7) |
| 290 | 0 | 530(15) | 4.2(1) | 51(6) | 151(8) |
| 291 | 0 | 635(16) | 3.1(1) | 43(5) | 164(10) |
| 292 | 0 | 725(15) | 4.0(1) | 15(2) | 278(19) |
| 293 | 298(27) | 637(29) | 2.7(2) | 9(1) | 359(20) |

### 3.2. Phase sequence and non-isothermal crystallization investigated by DSC scans

The DSC results (Figures 5 and 6, Table 2) show that 3F4HPhF6 crystallizes even for a cooling rate equal to 40 K/min, however, there are differences in the sample's behavior for slow and fast cooling/heating rates. Above the melting temperature of a crystal, the phase sequence is the same for all applied rates. The anomaly related to the Iso/SmC* transition has an onset at 380 K and corresponds to the absolute enthalpy change of 6 kJ/mol. The SmC*/SmC$_A$* transition occurs at 356 K during cooling and 360 K during heating. The absolute enthalpy change related to this transition is smaller than 0.1 kJ/mol. The exothermic anomaly related to crystallization shifts strongly towards lower temperatures with an increasing cooling rate. It is usual for the melt crystallization [30-33]. Using the onset temperature $T_o$ and the peak temperature $T_p$ of the crystallization-related anomaly determined for various cooling rates $\phi$, the activation energy $E_K$ (or $E_{AB}$) of the non-isothermal crystallization can be obtained by the Kissinger method [34]:

$$\ln\left(\frac{\phi}{T_p^2}\right) = C_K - \frac{E_K}{RT_p}, \qquad (4)$$

and the Augis-Bennett method [35]:

$$\ln\left(\frac{\phi}{T_p - T_o}\right) = C_{AB} - \frac{E_{AB}}{RT_p}. \qquad (5)$$

The activation plots for both methods are presented in Figure 7. The Kissinger method and the Augis-Bennett method lead to the same conclusions, and the effective activation energies are equal within uncertainties. The slope of the activation plot is larger for lower cooling rates, 5-15 K/min, than for higher ones, 20-40 K/min. The corresponding activation energies are $E_K = -174(10)$ kJ/mol and $E_{AB} = -170(10)$ kJ/mol for slow cooling and $E_K = -83(3)$ kJ/mol and $E_{AB} = -80(3)$ kJ/mol for fast cooling (positive slopes mean negative activation energies). In the latter case, the absolute value of the activation energy decreases by half. It means that in lower temperatures, where the molecular dynamics slow down, the diffusion rate has a larger influence on the crystallization kinetics than in higher temperatures. However, the influence of the diffusion rate is still smaller than the nucleation rate because the effective activation energies are negative. Different crystallization activation energies for slow and fast cooling are often reported for organic compounds [23,30,36-40].

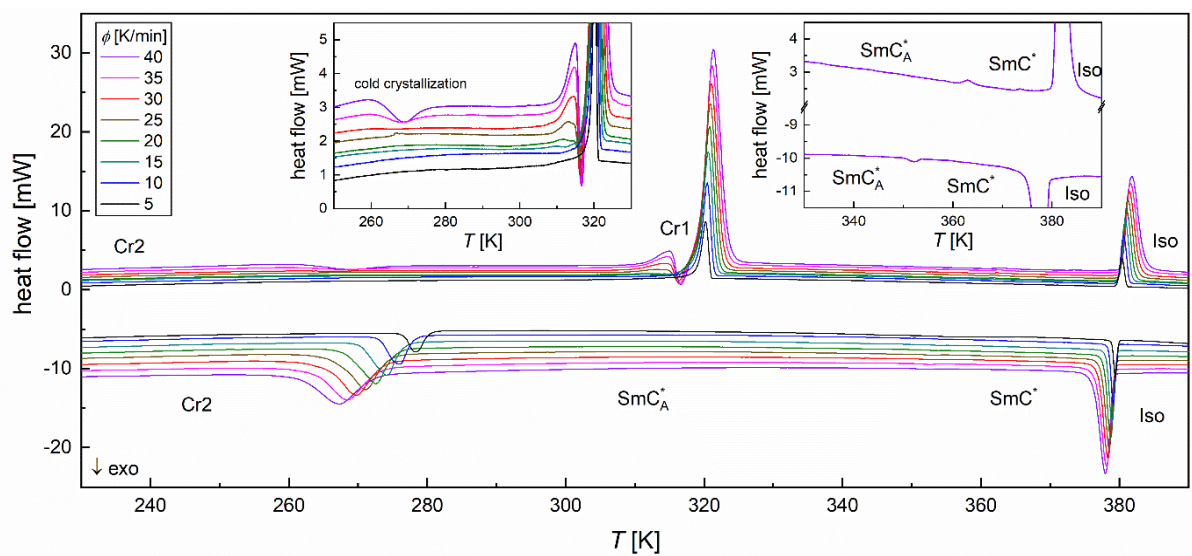

Figure 5. The DSC curves of 3F4HPhF6 collected for cooling/heating rates of 5-40 K/min. The insets show the magnified parts containing smaller anomalies related to cold crystallization and the SmC*/SmC$_A$* transition.

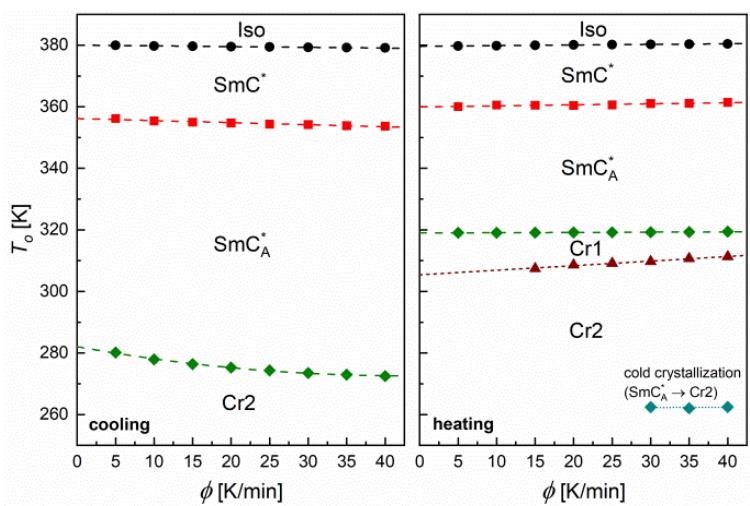

Figure 6. Onset temperatures of the phase transitions of 3F4HPhF6 obtained from the DSC curves collected for the 5-40 K/min rates during cooling (left panel) and heating (right panel).

Table 2. Onset $T_o$ and peak $T_p$ temperatures, enthalpy change $\Delta H$ and entropy change $\Delta S$ determined for the phase transitions of 3F4HPhF6. The values were obtained as a linear extrapolation to 0 K/min, if not stated otherwise.

| transition | $T_o$ [K] | $T_p$ [K] | $\Delta H$ [kJ/mol] | $\Delta S$ [J/(mol·K)] |
|---|---|---|---|---|
| Iso → SmC* | 380.0 | 379.6 | −6.1 | −16.0 |
| SmC* → SmC$_A$* | 356.2 | 355.4 | −0.05 | −0.15 |
| SmC$_A$* → Cr2 | 282.1[a] | 280.4[a] | −13.0[a] | −46.3[a] |
| SmC$_A$* → Cr2 | 262.5[b] | 268.8[b] | −1.2[b] | −4.3[b] |
| Cr2 → Cr1 | 305.4 | 306.5 | 1.4[b] | 4.6[b] |
| Cr1 → SmC$_A$* | 319.0 | 320.1 | 21.0 | 65.8[c] |
| SmC$_A$* → SmC* | 360.0 | 360.5 | 0.07 | 0.18 |
| SmC* → Iso | 379.7 | 380.2 | 6.3 | 16.6 |

[a] parabolic extrapolation instead of a linear one, [b] values for 40 K/min, [c] value for 5 K/min

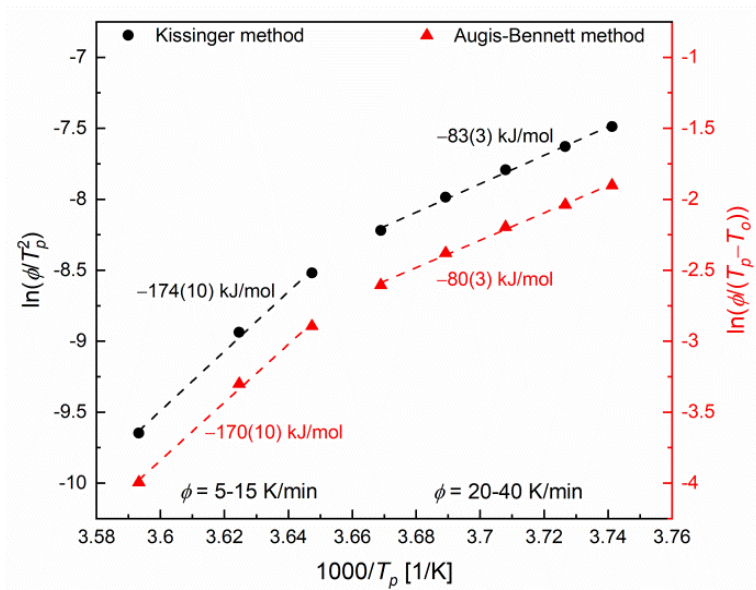

Figure 7. Activation plots to determine the effective activation energy of the non-isothermal crystallization of 3F4HPhF6 by the Kissinger method and the Augis-Bennett method (left and right axis, respectively).

The Augis-Bennett analysis also allows to determine the Avrami parameter using the formula [35]:

$$n = \frac{2.5RT_p^2}{E_{AB}\Delta T_{FWHM}}, \tag{6}$$

where $\Delta T_{FWHM}$ is the full-width at half-height of the exothermic anomaly on the DSC curve, corresponding to the crystallization. Two approaches were applied for 3F4HPhF6, where the activation energy differs for slow and fast cooling. In the first one, the $E_{AB}$ values determined from the linear fits, separately for 5-15 K/min and 20-40 K/min, were inserted into the (6) formula. In the second approach, the effective activation energy was obtained as a function of $\phi$ by calculation of the local slope of the Augis-Bennett plot. The second way gives less scattered $n$ values for 15 and 20 K/min, where the activation energy changes (Figure 8). In most cases, the Avrami parameter lies in the 3-4 range, corresponding to the isotropic growth of crystallites [25-27]. At this point, one can mention that the Kissinger method was developed for situations where $n = 1$ [34]. Despite that, it gives the activation energy consistent with the results of the Augis-Bennett method, which is also applicable for $n > 1$ [35].

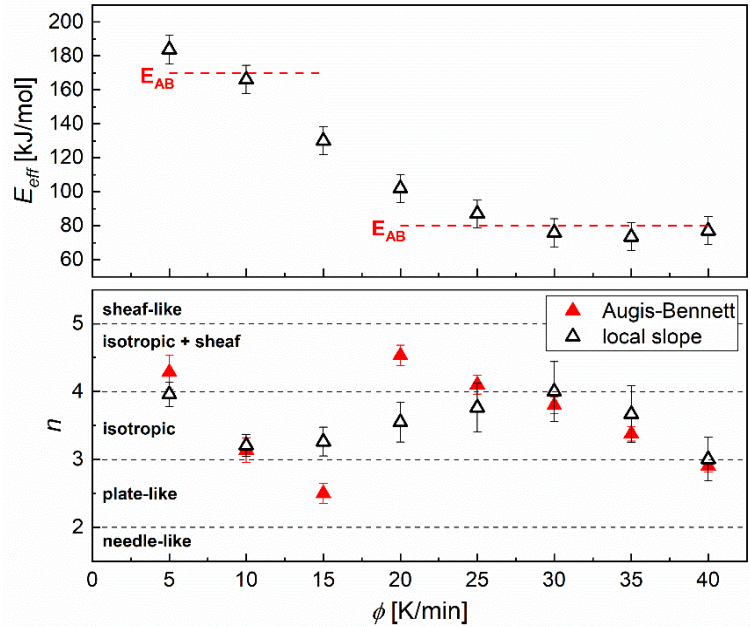

Figure 8. Effective activation energy determined from the local slope of the Augis-Bennett plot and the Avrami exponent obtained from the (6) formula vs. cooling rate. The interpretation of the $n$ values is based on [25-27].

The absolute enthalpy change related to crystallization decreases with an increasing cooling rate (Figure 9), which is interpreted as incomplete crystallization. It is confirmed by the small exothermic anomaly observed during heating at ca. 270 K, which indicates the cold crystallization of the remaining SmC$_A$* phase. The change of $\Delta H$ with $\phi$ for the SmC$_A$* → Cr2 transition is significant enough to make an extrapolation to $\Delta H = 0$ and to estimate the cooling rate enabling complete avoidance of melt crystallization. The simple linear extrapolation indicates the threshold cooling rate of 81 K/min, while the parabolic fit gives 64 K/min. Using $\Delta H = -13.0$ kJ/mol extrapolated by the parabolic fit to 0 K/min, one can estimate that for the 40 K/min cooling rate, the degree of melt crystallization is 52.5 % and almost half of the sample remains in the SmC$_A$* phase.

During heating at 15-40 K/min, a small anomaly indicating the transition between two crystal phases, Cr2 and Cr1, is observed at 307-311 K (Figures 5 and 6). The melting temperature is 319 K regardless of the cooling/heating rate, therefore the final phase before melting is always Cr1. The Cr2 → Cr1 transition is not noticed for the 5 and 10 K/min rates, probably because it starts at lower temperatures and the wide and small anomaly is not visible over the baseline. The melting temperature is 9 K lower than 328 K reported in Ref. [2], which means that the crystal phase formed from the melt has a different structure than the crystal phase of the pristine sample.

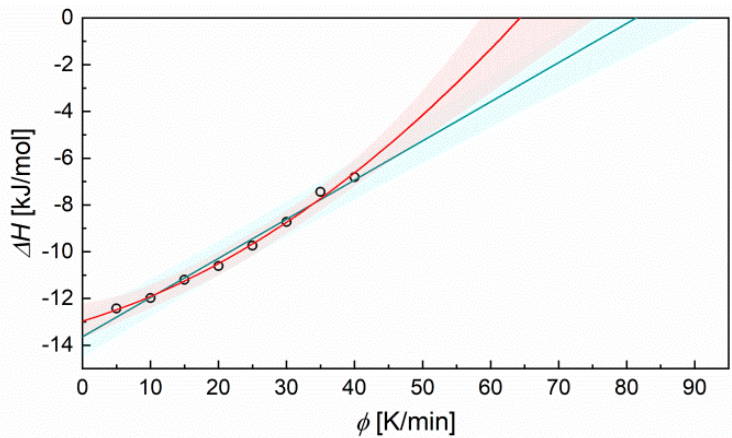

Figure 9. The enthalpy change during crystallization plotted against the cooling rate with the linear and parabolic extrapolations to $\Delta H = 0$ and $\phi = 0$. Light red and green bands around the fitted curves are the 95% certainty regions.

### *3.3. Structural analysis based on X-ray diffraction patterns*

The XRD patterns presented in Figure 10 were collected for polycrystalline 3F4HPhF6 in three temperature programs:
- Figure 10a: the pristine sample (not previously melted after synthesis) heated up to 343 K,
- Figure 10b: the sample heated to 343 K to the SmC$_A$* phase, cooled down to 268 K at the 5 K/min rate, and subsequently heated back to 343 K,
- Figure 10c: the sample heated to 343 K to the SmC$_A$* phase, cooled down to 303 K at the 5 K/min rate, and after five hours subsequently heated back to 343 K.

The characteristic period $d$, determined using the Bragg equation [41] from the positions of the low-angle 1$^{st}$ order peak and corresponding peaks of higher orders (Figure S2 in SM), is presented in Figure 10d. The XRD pattern of the crystal phase of the pristine sample (Figure 10a) contains a strong low-angle diffraction peak at $2\theta \approx 2.8°$ and the $d$ period equals on average 31.65(3) Å. The diffraction peak at $2\theta \approx 3.0°$ indicates the lamellar structure of the SmC$_A$* phase [41,42], and the smectic layer spacing is equal to 29.49(2) Å. The decrease of $d$ upon melting of the pristine sample at 331 K is 7.2(1)%, where 100% is the smectic layer spacing.

After the sample is cooled down with the 5 K/min rate to 268 K, it is already crystallized (Figure 10b). The low-angle peak is observed at $2\theta \approx 2.6°$, at a different position than for the pristine sample. The peaks at higher $2\theta$ angles are wide and overlapped, indicating a large number of defects. The detailed analysis of the low-angle peak's position shows that at temperatures of 268-293 K, the average period is 33.70(2) Å, and in 303-320 K, it increases to 34.12(2) Å. The small jump of $d$ between 293 and 303 K corresponds to the Cr2 → Cr1 transition, which was visible on the DSC curves at 307-310 K for higher heating rates and was presumed to occur at lower temperatures during slow heating. The crystal phase melts above 320 K, with a 15.9(1)% decrease of $d$.

After cooling to 303 K (Figure 10c), the sample also crystallizes in the crystal phase with the low-angle diffraction peak at $2\theta \approx 2.6°$ and the characteristic period of 34.05(7) Å, larger than in the pristine sample, and the decrease of $d$ during melting is 15.6(4)%. In the first pattern collected after cooling down to 303 K, the sample is only partially crystallized, and crystallization is completed in the third pattern. The time of collection of one XRD pattern was 17 min, after 5 min of temperature stabilization, therefore, the isothermal crystallization at 303 K lasts less than one hour. The pattern of the crystal phase formed at 303 K consists of sharper peaks than after cooling down to 268 K. The POM observations show that the size of crystallites increases with increasing crystallization temperature. The XRD results also confirm that the pristine sample consists of a crystal phase with a higher melting temperature than the crystal phase formed from the melt.

The length of the 3F4HPhF6 molecule, calculated by the DFT method, is 34-37 Å, depending on conformation [43]. Based on this, one can presume that both in the pristine sample and after crystallization from the melt, the crystal phases, with the characteristic periods respectively of ca. 31.7 and 33.7-34.1 Å, have the lamellar structure, which is often observed also for other smectogens in the crystal state [44-46]. In the pristine sample, the molecules are likely tilted in the layers, although by a smaller angle than in the SmC$_A$* phase, while in the crystal phase formed from the melt, the long axes of molecules are rather orthogonal to the layer plane. As our XRD setup does not provide high cooling rates, this method cannot investigate the partial crystallization accompanied by vitrification of the SmC$_A$* phase.

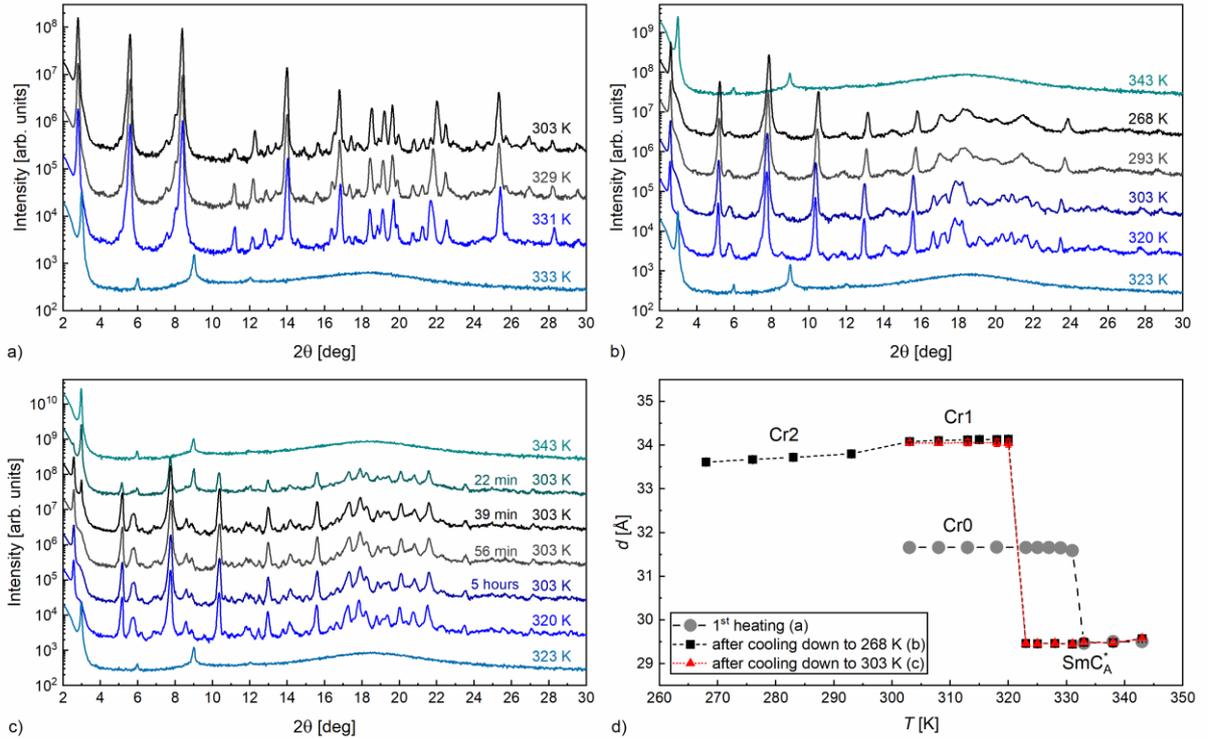

Figure 10. Selected XRD patterns of 3F4HPhF6 collected during three temperature programs: the 1$^{st}$ heating of the pristine sample (a), cooling directly from 343 K to 268 K and subsequent slow heating to 343 K (b) and cooling directly from 343 K to 303 K, isothermal measurement at 303 K for five hours and slow heating to 343 K (c). In each panel (a-c), the patterns are shown chronologically from top to bottom. The characteristic period in the crystal and smectic phases is shown in (d) as a function of increasing temperature. The lines in (d) are guides for the eye.

*3.4. Dielectric relaxation processes studied by BDS*

The 3F4HPhF6 sample was heated to the isotropic liquid and cooled down to 213 K with the 15 K/min rate. The dielectric spectra collected subsequently upon heating are presented in Figure S3 as a map of the dielectric absorption vs. frequency and temperature. The spectra were fitted using the Cole-Cole model [47,48]:

$$\varepsilon^*(f) = \varepsilon'(f) - i\varepsilon''(f) = \varepsilon_\infty + \sum_j \frac{\Delta\varepsilon_j}{1+(2\pi i f \tau)^{1-a}} - \frac{i\sigma}{2\pi\varepsilon_0 f}, \quad (7)$$

where $\varepsilon^*$ is the complex dielectric permittivity of the sample, consisting of the dispersion (real) part $\varepsilon'$ and the absorption (imaginary) part $\varepsilon''$, $f$ is frequency, $\varepsilon_\infty$ is dielectric dispersion in the high frequency limit, $\Delta\varepsilon$ is dielectric strength, $\tau$ is relaxation time, $a$ is the parameter of the relaxation time distribution, $\sigma$ is ionic conductivity and $\varepsilon_0 = 8.85 \cdot 10^{-12}$ F/m is the vacuum permittivity. If $a = 0$, one obtains the Debye model. The Cole-Cole plots with fitted Equation (7) for different relaxation processes are presented in Figure S4, while Figure 11 shows the obtained temperature dependence of dielectric strength, relaxation time and ionic conductivity.

The BDS spectra contain two very weak relaxation processes in the crystal phase. The high-frequency process is visible up to 220 K. It cannot be analyzed by the Cole-Cole model because it overlaps with artifacts above 1 MHz. The low-frequency process, denoted as the cr-process, is observed above 240 K. It is described by the Cole-Cole model with the parameter $a \approx 0.3$, and its relaxation time changes with temperature according to the Arrhenius formula. The activation energy of this process is 67.4(6) kJ/mol. The presence of the relaxation process in the crystal phase indicates that it is either the conformationally disordered (CONDIS) or orientationally disordered (ODIC) phase [49,50]. The DFT calculations for 3F5HPhF6 (B3LYP exchange-correlation functional, TZVPP basis set), which differs from 3F4HPhF6 by one additional $CH_2$ group in the non-chiral chain, give the energy barrier of 35 kJ/mol for the rotation of a fluorinated ring and 35 kJ/mol for the rotation of a biphenyl within the aromatic core of molecules [11], which sums to the experimental activation energy of the cr-process. As both for m = 4 and 5, the structure of the molecular core is the same, the energy barriers obtained for m = 5 can be applied also to m = 4. Thus, the cr-process observed in the crystal phase of 3F4HPhF6 may be attributed to the intra-molecular rotations of a biphenyl and fluorinated ring. In such a case, 3F4HPhF6 forms the CONDIS phase.

In the SmC$_A$* phase, there are two relaxation processes: P$_L$ phason at lower frequencies and P$_H$ phason at higher frequencies (both with $a \approx 0.1$), which are the collective fluctuations around the tilt cone, in-phase in neighbor smectic layers for P$_L$ and anti-phase for P$_H$ [51,52]. The P$_L$ process shows the Arrhenius dependence, probably because it overlaps with the molecular s-process (rotations around the short molecular axis) [51], and the activation energy is 106.5(7) kJ/mol, while the P$_H$ relaxation time changes in the non-Arrhenius manner. In the SmC* phase, the strong Goldstone mode ($a \approx 0.3$) is related to oscillations around the tilt cone [53-55]. The decrease of the dielectric strength and relaxation time of the Goldstone mode with increasing temperature is probably caused by decreasing helical pitch [55], although the exact temperature dependence of the pitch in 3F4HPhF6 is not known because its values exceed 1.4 μm, which is a limit for the transmission method [3]. The detailed investigation of relaxation processes in the SmC$_A$* and SmC* phases in the external bias field has been published in Ref. [6] for a thinner sample (5 μm). The BDS spectra registered in the isotropic liquid phase contain one Debye relaxation process ($a = 0$) at very high frequencies, with the activation energy equal to 61(1) kJ/mol. It can be either the molecular s-process or l-process (rotations around the long molecular axis). The more probable is the s-process because if it were the l-process, one would observe a slower s-process at lower frequencies [50].

The conductivity often shows discontinuous changes at the phase transitions [50,56]. The $\sigma$ values of 3F4HPhF6 increase with increasing temperature in the Arrhenius manner, although the activation energy depends on a particular phase. The largest activation energy equal to 105(2) kJ/mol is obtained for the crystal phase. In the SmC$_A$* phase, it is reduced to 52(1) kJ/mol, and in the SmC* phase, it is the smallest and equal to 12(5) kJ/mol. After the transition to the isotropic liquid phase, the activation energy increases to 75(2) kJ/mol. Each phase transition from the sequence: crystal → SmC$_A$* → SmC* → Iso, leads to a discontinuous increase in the ionic conductivity. In the crystal phase below 243 K, the ionic conductivity was too small to be determined from the registered BDS spectra.

The vitrified SmC$_A$* phase was not detected by the BDS method. Based on the DSC results, the fraction of the remaining SmC$_A$* phase after melt crystallization during cooling at 15 K/min is only 14%, which is apparently not enough for observation of the α-process typical for the supercooled liquid/liquid crystal and the secondary β-process typical for the glassy state [5-7,11,23,57,58]. However, results for 3F5HPhF6 and 3F6HPhF6 presented in [7] show that the relaxation time of the $P_H$ phason follows the Vogel-Fulcher-Tammann formula $\tau(T) = \tau_0 \exp(B/(T - T_V))$ similarly as the α-relaxation time, and that both processes slow down significantly in the same temperature region, i.e. when approaching the glass transition. For 3F4HPhF6, fitting the VFT formula to the α-relaxation time cannot be performed because the sample quickly crystallizes in the temperature region where the α-process enters the achievable frequency range. However, it can be done instead for the $P_H$ phason to estimate the glass transition temperature. The fitting, shown in Figure 11(b), is done for the relaxation times determined on heating and additionally for two points obtained after cooling at 15 K/min from isotropic liquid directly to 293 K and 283 K. The fitting parameters are as follows: $\log_{10} \tau_0 = -9.3(2)$, $B = 771(76)$ K, $T_V = 214(6)$ K. If the glass transition temperature is defined as the one where $\tau = 100$ s [58], then $T_g = 244(3)$ K. The slowing down of the molecular dynamics, which is significant enough to hinder crystallization, occurs at even higher temperatures: on the DSC curves, both melt and cold crystallization are observed only above 260 K.

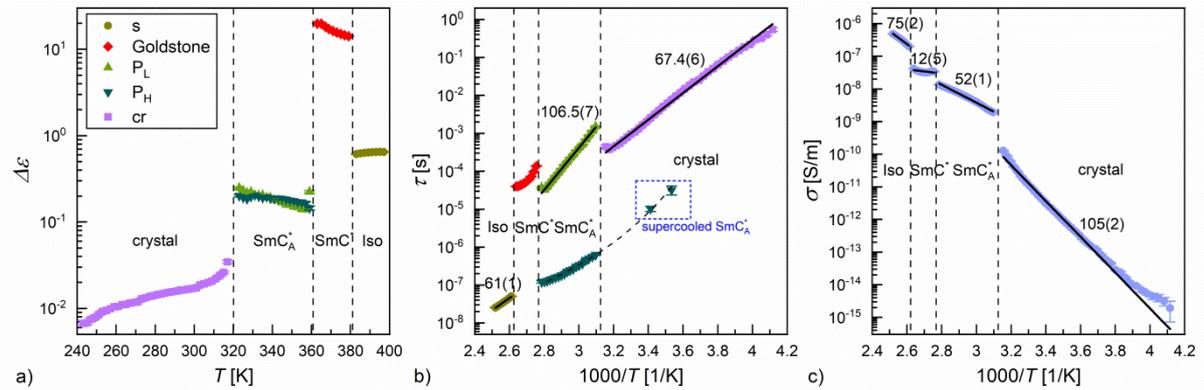

Figure 11. BDS results for 3F4HPhF6: dielectric strength as a function of temperature (a) and activation plots of relaxation time (b) and ionic conductivity (c). Panels (a) and (b) have a common legend. The activation energies are given in kJ/mol. The results were obtained upon heating a sample cooled from isotropic liquid with the 15 K/min rate, except for two points in a frame in (b), each obtained after cooling at 15 K/min from isotropic liquid.

## 4. Summary and conclusions

The 3F4HPhF6 compound undergoes melt crystallization at 5-40 K/min cooling rates. The final degree of crystallization decreases with an increasing cooling rate. The significant slowing down of the molecular mobility, preventing further crystallization, occurs below 260 K, and the estimated dynamic glass transition temperature of the SmC$_A$* phase is 244 K. The predicted cooling rate necessary to obtain the pure glass of the SmC$_A$* phase is quite high, at least 64 K/min. Thus, from a practical point of view, 3F4HPhF6 is not the best candidate to be used to prepare the SmC$_A$* glass, and the homologues with m = 5, 6, 7 are more suitable for that [5,6]. On the other hand, 3F4HPhF6 can be tested as a component of glassforming mixtures consisting of antiferroelectric liquid crystals from the 3FmHPhF6 and similar series, as this subject (the vitrification of such mixtures) has not been widely investigated [11].

Three crystal phases were observed for 3F4HPhF6, all apparently with the lamellar structure resembling the structure of a smectic phase:

- The Cr0 phase in the pristine sample, which is characterized by the layer spacing of 31.7 Å. According to the DSC results from Ref. [2], Cr0 melts at $T_m = 328.1$ K with $\Delta H_m = 28.4$ kJ/mol and $\Delta S_m = 86.6$ J/(mol·K). Formation of Cr0 from the melt was not observed.

- The Cr2 phase formed during crystallization from the melt below ca. 300 K, characterized by the layer spacing of 33.7 Å and transforming on heating to the Cr1 phase above 300 K with $\Delta H$ = 1.4 kJ/mol and $\Delta S$ = 4.6 J/(mol·K).
- The Cr1 phase formed by the Cr2 → Cr1 transition on heating or during crystallization from the melt above 300 K. The layer spacing in Cr1 equals 34.1 Å. Cr1 melts at $T_m$ = 319.0 K with $\Delta H_m$ = 21.0 kJ/mol and $\Delta S_m$ = 65.8 J/(mol·K). According to the XRD patterns, Cr2 and Cr1 have very similar structures, which agrees with a small energy effect of the transition between them. As the BDS spectra reveal, both Cr2 and Cr1 are the CONDIS phases. The conformational disorder is probably related to rotations of the aromatic rings in the molecular core and shows in the BDS spectra as a weak relaxation process with an activation energy of 67 kJ/mol.

The role of the translational degrees of freedom was indicated in the previous works regarding melt crystallization in similar compounds [59,60]. Molecules in the SmC$_A$* phase have two translational degrees of freedom, corresponding to the movements within the smectic layer, and the thermal energy for each degree of freedom is $RT/2$. As it was noticed in [59,60], the intersection of the thermal energy of one or two translational degrees of freedom with the $\Delta S_m(T_m - T)$ function [29], which describes the thermodynamic driving force of crystallization, may correspond to a temperature or temperature range characteristic for crystallization of a given compound. The plot for 3F4HPhF6 is presented in Figure 12. The melt crystallization occurs in the Cr2 or Cr1 phase. Due to the Cr2 → Cr1 transition, the $\Delta S_m(T_m - T)$ function can be calculated only for Cr1. As Cr2 and Cr1 both are the CONDIS phases with similar crystal structures, it is assumed that the $\Delta S_m^{Cr1}(T_m^{Cr1} - T)$ function can be applied also for crystallization below 303 K. The intersection temperature of $RT$ and $\Delta S_m^{Cr1}(T_m^{Cr1} - T)$ is 283 K, while the onset temperature of crystallization determined by DSC is just below, in the 273-280 K range. Moreover, the intersection temperature of $RT/2$ and $\Delta S_m^{Cr1}(T_m^{Cr1} - T)$ is 300 K, above which the Cr2 phase transforms into Cr1 (full Cr2 → Cr1 transition observed in 303 K in the XRD patterns, onset temperature 311.3 K for 40 K/min obtained by DSC). We are aware that the observed correlations between the thermal energy and thermodynamic driving force may be accidental, but there is a plan to check these relationships for a larger number of compounds, and we would also like to attract the attention of other researches into looking for such connections in crystallization studies. The thermodynamic driving force of Cr0 is larger than for Cr1, also, the layer spacing is closer to the one in the SmC$_A$* phase (29.5 Å) than for Cr2 and Cr1. Despite that, Cr0 is not formed from the melt. A possible reason is that the intra-layer arrangement of molecules in Cr2 and Cr1 could be closer to the one in the SmC$_A$* phase, contributing to their easier formation. Above the melting temperature of Cr1, the $\Delta S_m^{Cr0}(T_m^{Cr0} - T)$ function is smaller than $RT/2$, thus the crystallization to the Cr0 phase in the 319-328 K range is very unlikely.

The POM observation of the crystallization process enabled the determination of the activation energy for nucleation equal to 409 kJ/mol, which means that the number of crystallites strongly increases with decreasing temperature. This causes a decrease in the average size of crystallites in the range of 110-360 μm for 298-293 K, but less than 1 μm when extrapolated to 270 K. As 3F4HPhF6 can be supercooled tenths of a degree below the melting temperature, it gives the opportunity to prepare composites which consists of the Cr2 crystallites with a tunable size mixed with the SmC$_A$* phase or the SmC$_A$* glass. Experiments with a sample additionally aligned by the electric field may be particularly interesting.

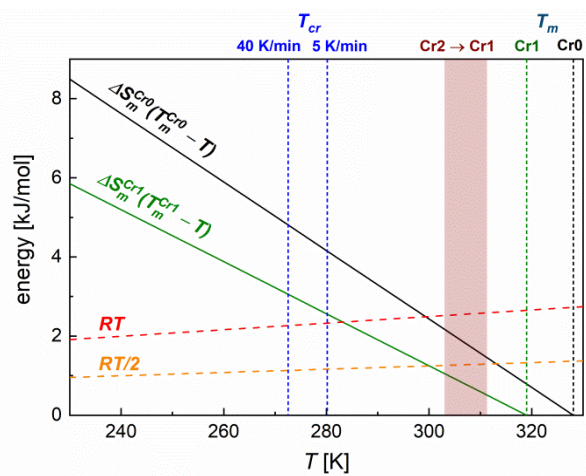

Figure 12. Comparison of the thermodynamic driving force of crystallization of 3F4HPhF6 with the thermal energy of translational degrees of freedom.

**Authors' contribution:**
A. Deptuch – conceptualization, investigation (POM, XRD), formal analysis (POM, DSC, XRD, BDS), writing – original draft
S. Lalik – investigation (DSC), formal analysis (DSC), writing – review and editing
M. Jasiurkowska-Delaporte – investigation (BDS), writing – review and editing
M. Urbańska – resources (sample's synthesis), writing – review and editing
M. Marzec – funding acquisition, resources, writing – review and editing

**Acknowledgments:** The PerkinElmer DSC 8000 calorimeter was purchased thanks to the financial support of the European Regional Development Fund in the framework of the Polish Innovation Economy Operational Program (contract no. POIG.02.01.00-12-023/08). This research was supported in part by the Excellence Initiative – Research University Program at the Jagiellonian University in Kraków.

**Conflicts of interest statement:** There are no conflicts to declare.

# Crystallization kinetics of (S)-4'-(1-methylheptyloxycarbonyl)biphenyl-4-yl 4-[4-(2,2,3,3,4,4,4-heptafluorobutoxy)but-1-oxy]-2-fluorobenzoate

Aleksandra Deptuch[1,*], Sebastian Lalik[2], Małgorzata Jasiurkowska-Delaporte[1], Magdalena Urbańska[3], Monika Marzec[2]

[1] Institute of Nuclear Physics Polish Academy of Sciences, Radzikowskiego 152, PL-31342 Kraków, Poland
[2] Institute of Physics, Jagiellonian University, Łojasiewicza 11, PL-30348 Kraków, Poland
[3] Institute of Chemistry, Military University of Technology, Kaliskiego 2, PL-00908 Warsaw, Poland
[*] corresponding author, aleksandra.deptuch@ifj.edu.pl

# Supplementary Materials

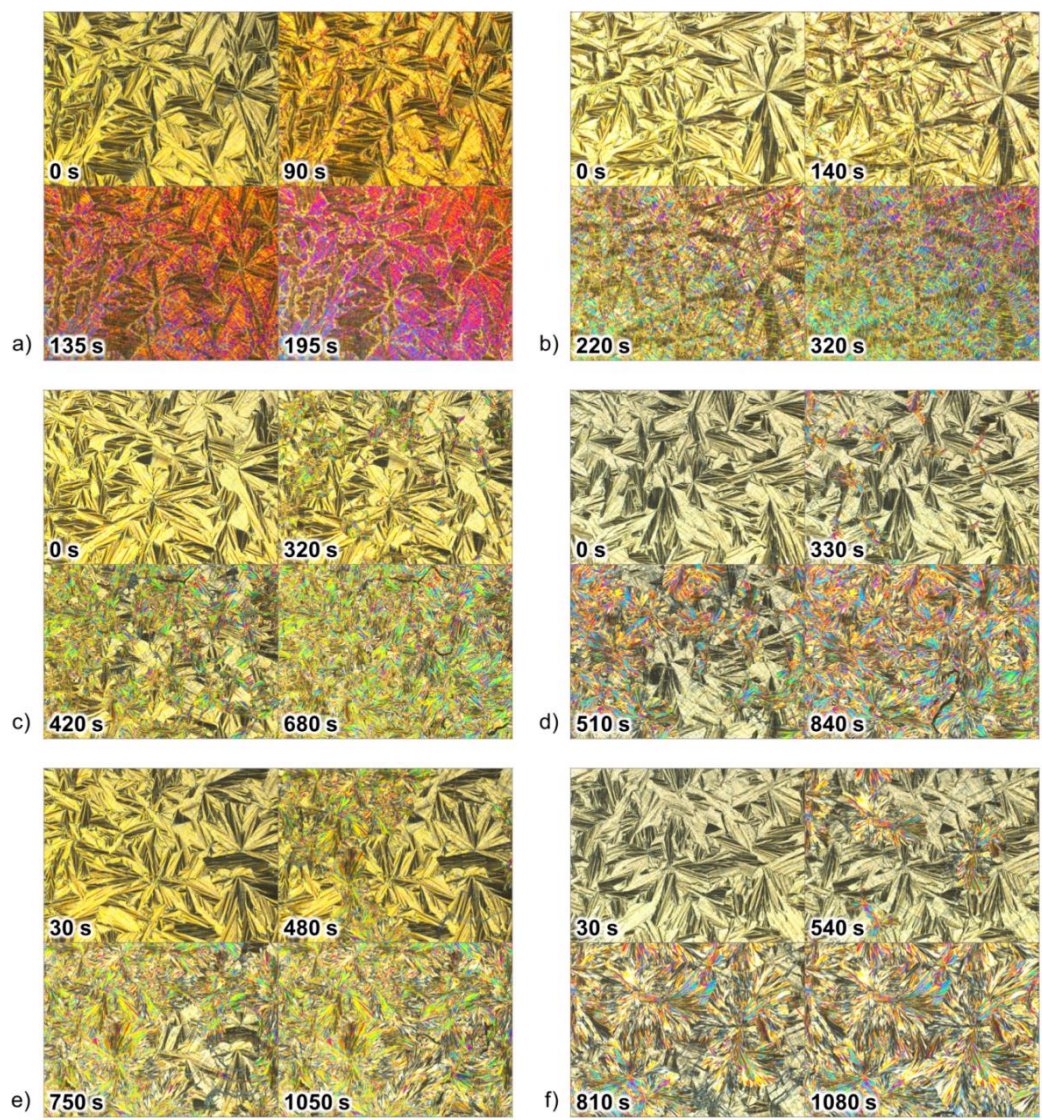

Figure S1. POM textures registered during isothermal melt crystallization in 285 K (a), 287 K (b), 289 K (c), 290 K (d), 291 K (e), and 292 K (f) after cooling with the 30 K/min rate from the isotropic liquid phase.

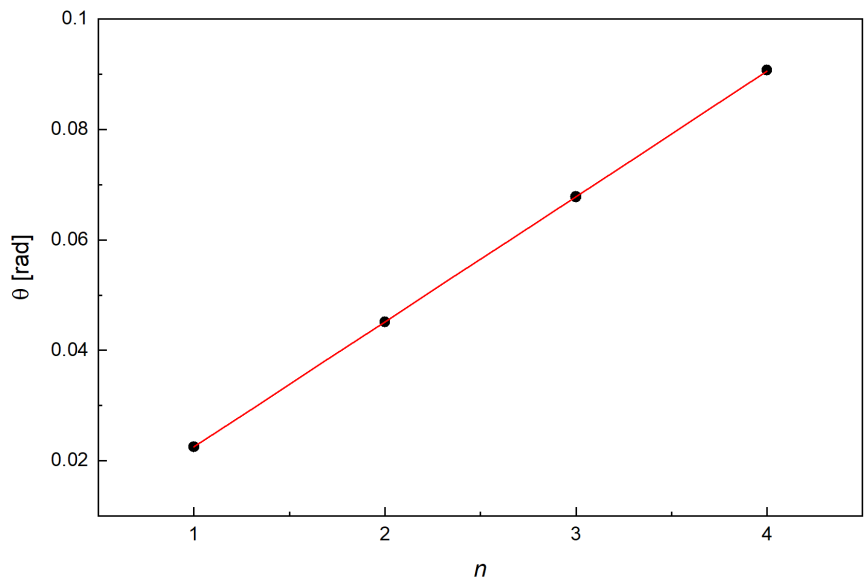

Figure S2. An example of determination of the characteristic period $d$ for a sample crystallized in 303 K. The fitted function is $\theta(n) = \theta_0 + \arcsin(l\lambda/2d)$, where $\theta$ are positions of the $l^{th}$ order diffraction peaks, $\theta_0$ is the systematic shift in the peak's position and $\lambda = 1.5406$ Å is the CuKα radiation wavelength. The function is another form of the Bragg equation $l\lambda = 2d\sin(\theta - \theta_0)$.

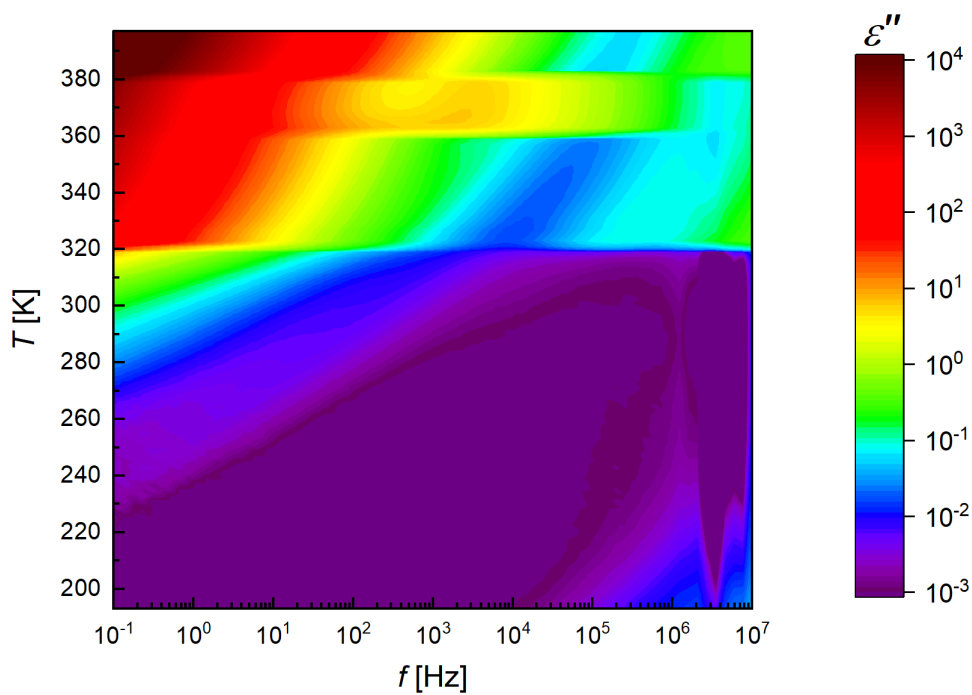

Figure S3. Dielectric absorption of 3F4HPhF6 upon heating after cooling from the isotropic liquid with the 15 K/min rate.

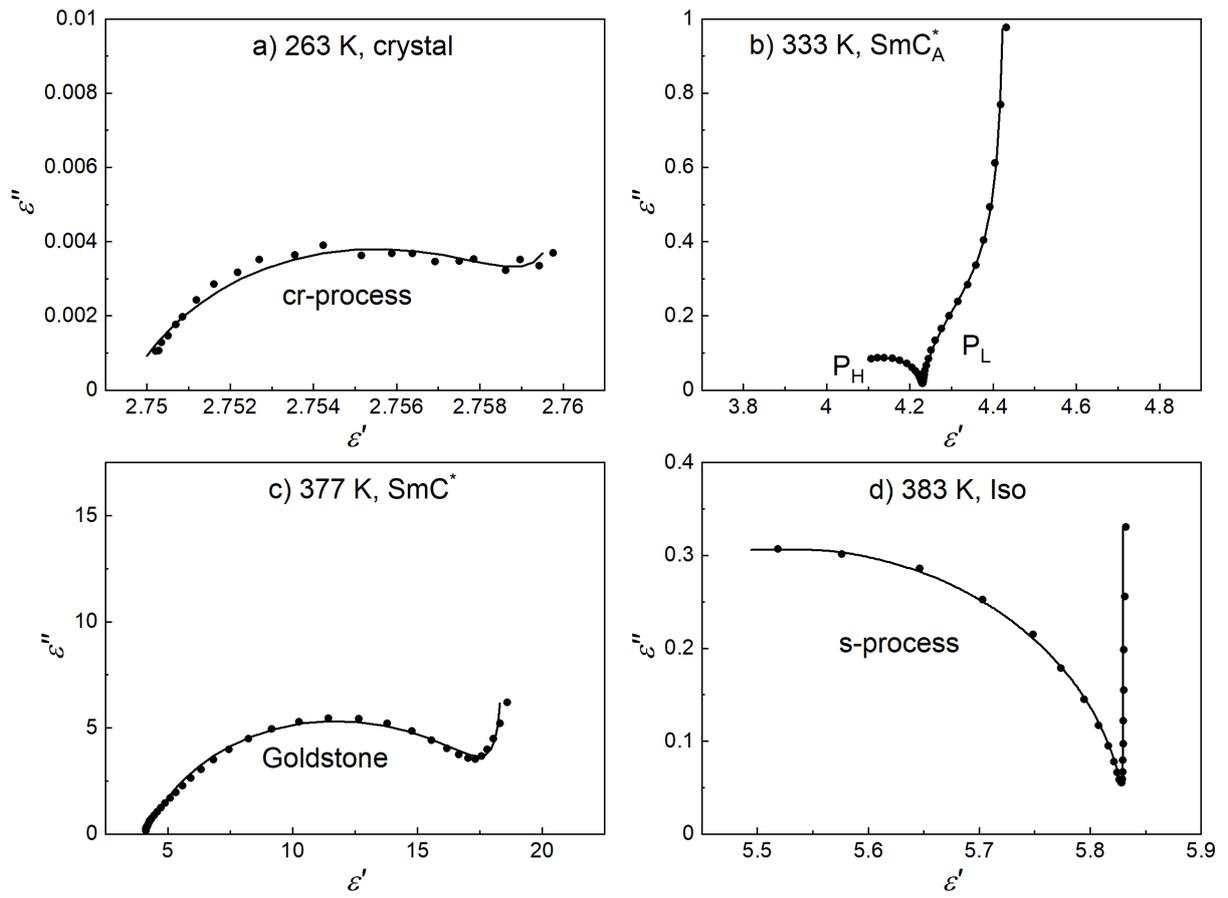

Figure S4. Cole-Cole plots of experimental BDS spectra (points) and fitting results of Equation (1) (lines) for different phases of 3F4HPhF6.